# Transmission-Mode Silicon-Rich Nitride Mie-Void Metasurfaces in the Visible

*Oren Goldberg\*, Noa Mazurski and Uriel Levy*

*Oren Goldberg[1], Noa Mazurski[1], and Uriel Levy[1,2]*

*1 - Institute of Applied Physics, The Faculty of Science, The Center for Nanoscience and Nanotechnology, The Hebrew University of Jerusalem, Jerusalem 91904, Israel*

*2 - Singapore-HUJ Alliance for Research and Enterprise (SHARE), The Smart Grippers for Soft Robotics (SGSR) Programme, Campus for Research Excellence and Technological Enterprise (CREATE), Singapore 138602, Singapore*

*Corresponding author: Oren.goldberg@mail.huji.ac.il*

Keywords: transmission, Mie voids, metasurfaces, structural colors

**Abstract**

Mie-void metasurfaces have so far been developed mainly in reflection, where subwavelength voids embedded in high-index media support localized resonances and spectrally selective optical responses. Yet, many optical systems could benefit from integrating such optical elements operating in transmission mode. Motivated by this great need, we hereby introduce Mie-void metasurfaces operating in transmission. To allow for their operation in the visible range, our Mie-voids are implemented using the silicon-rich nitride (SRN) platform. We show that this transition from reflection to transmission is not a simple change in geometry: placing the voids in a finite film on a substrate introduces slab-guided and Fabry–Pérot-like contributions that hybridize with the underlying Mie-void response. Rigorous coupled-wave analysis shows that the dominant spectral transformation occurs when the semi-infinite host is replaced by a finite SRN film, while the substrate acts mainly as a secondary perturbation. Thickness-dependent dispersion maps reveal an avoided crossing between interacting modes, supporting the interpretation of a hybrid transmission regime and identifying film thickness as a clean parameter for tracking the evolution of the coupled modal structure. Experimentally, we realize transmission-mode structural colors by varying the void depth and observe good agreement between measured and simulated spectra and chromaticity coordinates. By spatially programming the void depth, we further demonstrate transmitted-light patterns and image encoding within a single metasurface architecture. These results establish transmission-mode Mie-void metasurfaces as a viable inverse-dielectric platform operating in transmission, with plethora of potential important applications such as transmissive spectral filtering, optical encoding, and display-oriented photonic elements, to name a few.

**Introduction**

Visible metasurfaces have emerged as compact platforms for shaping optical spectra and wavefronts[1–3] for applications such as filtering[4–6], imaging[7–9], and structural color generation[5,10–14]. In the visible regime, transmissive implementations are especially attractive because they are naturally compatible with optical systems that operate with forward-propagating light[7,15–18], including compact spectral filters[19–21], image-encoding elements[22,23], and display-oriented components[24–27]. Despite this relevance, many visible dielectric metasurfaces continue to emphasize reflective operation[5,28,29], while transmissive designs often rely on conventional positive resonator geometries[8,18,30,31]. As a result, inverse-dielectric metasurface architectures remain relatively unexplored in the visible transmission regime[32,33].

Mie-void metasurfaces provide an inverse-dielectric route to resonant light manipulation, in which subwavelength voids embedded within a high-index medium support spectrally selective optical response. Earlier work established this platform primarily in reflection, where it enabled visible spectral tuning and structural color generation. A central aspect of the Mie-void concept is that the optical response is governed mainly by the local void geometry, while not being fundamentally restricted to purely collective periodic effects[32]. This makes Mie-void metasurfaces a distinct resonant platform for extending inverse-dielectric photonics into transmissive operating geometries.

Translating the Mie-void platform from reflection to transmission is not a simple change in operating geometry. In reflection-mode implementations, the spectral response can often be understood largely in terms of resonances localized around the void itself. In transmission, however, the situation becomes more complex: a finite high-index dielectric film on a substrate can also support slab-guided[34–36] and Fabry–Pérot-like modes[37–40] that interact with the underlying Mie-void response. The key question is therefore whether the Mie-void concept remains a useful and designable framework in this transmission geometry, or whether the additional modal structure overwhelms the localized resonant picture that defines the platform.

Silicon-rich nitride (SRN) provides a suitable material platform for examining this transition at visible wavelengths, combining a high refractive index with sufficiently low optical loss in a nanofabricated thin-film geometry[41–44]. In this setting, the finite SRN layer is not merely a passive host for the voids, but an active part of the optical system that can mediate coupling between localized Mie-void resonances and film-supported modes. Within this framework, film thickness serves as a clean parameter for tracking how the transmission resonances evolve across the resulting hybrid regime.

In this work, we introduce visible transmission as an operating regime for Mie-void metasurfaces using an SRN platform. We show that, in contrast to the more localized resonant picture associated with reflection-mode Mie voids, transmission through a finite film on a substrate leads to a hybrid optical response involving Mie-void, slab-guided, and Fabry–Pérot-like contributions. Using film thickness as a controlled parameter, we track the evolution of these coupled resonances and identify the finite-film modal structure that governs the transmission response. Experimentally, we realize transmission-mode structural colors and spatially encoded patterns through depth-programmed fabrication, confirming that the predicted structured spectral behavior is retained in fabricated devices despite the added modal complexity. In parallel, grayscale nanolithography provides a practical route for continuously controlling void depth[45], enabling experimental access to this transmission regime. Together, these results establish transmission-mode Mie-void metasurfaces as a viable inverse-dielectric platform for transmissive spectral filtering, image encoding, and display-oriented optical components.

**Results**

**2.1 Establishing transmission-mode Mie-void metasurfaces in finite SRN films**

We begin from the canonical Mie-void picture of a cylindrical void embedded in a semi-infinite SRN medium, as illustrated schematically in Fig. 1a. In this idealized limit, the optical response is dominated by a resonance localized around the void, with its spectral position governed primarily by the lateral size and depth. This serves as the reference picture for conventional reflection-mode Mie-void platforms, in which the void acts as the principal resonant building block.

The experimentally relevant transmission-mode structure considered here is fundamentally different. As illustrated in Fig. 1b, the void is realized as part of a periodic array etched into a finite-thickness SRN film on a $SiO_2$ substrate, so that the transmitted response is shaped not only by the localized void resonance but also by the surrounding finite-film environment. In this geometry, the Mie-void response hybridizes with film-supported modes, giving rise to multiple coupled spectral features rather than a single isolated resonance. The central issue is therefore not simply whether Mie-void resonances remain observable in transmission, but whether the original Mie-void design logic remains meaningful in a finite-film system where localized and film-mediated contributions coexist. This question motivates the modal analysis developed below.

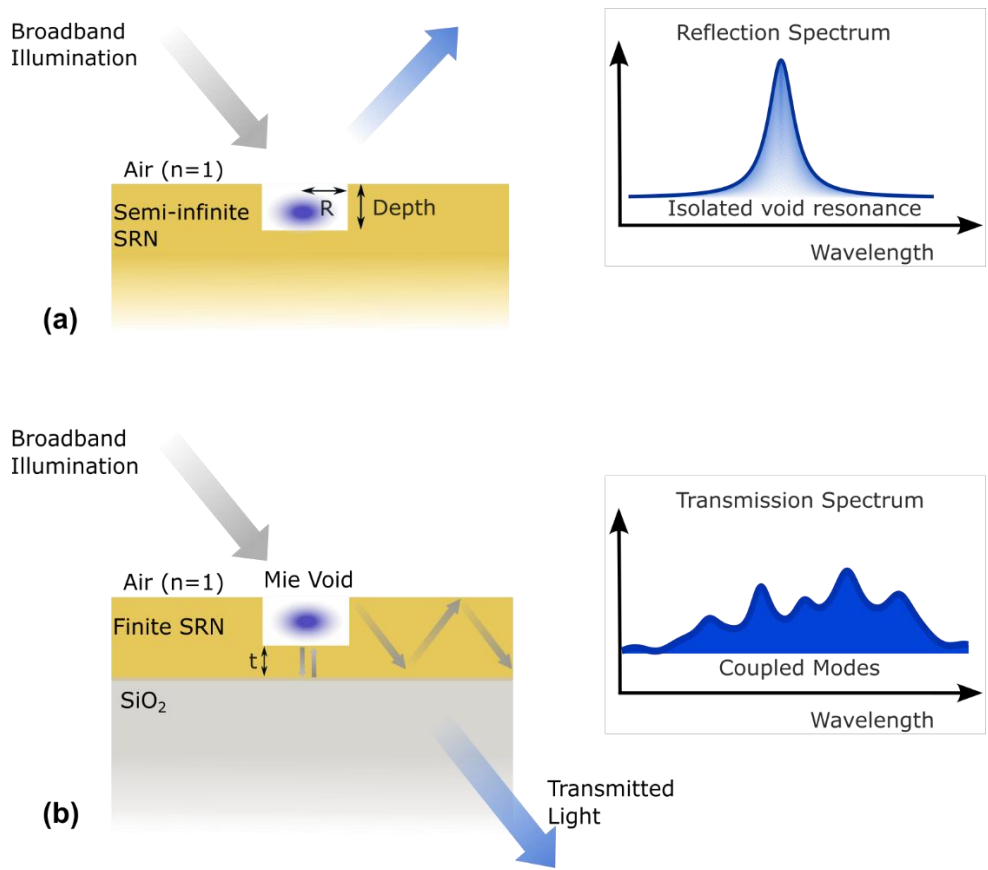

*Figure 1: a) Schematics of an isolated cylindrical void in a semi-infinite medium of SRN under broadband illumination. The void supports a localized Mie-type resonance characterized by its radius R and its depth, giving rise to a spectrally selective optical response. In this idealized case, the response can be viewed as that of an isolated void resonance. b) Schematics of the experimentally relevant structure considered in this work: a periodic array of cylindrical voids etched into a finite-thickness SRN film on a SiO₂ substrate. In transmission, the localized void resonances hybridize with the finite-film response, resulting in multiple coupled spectral features that can be tuned through the metasurface geometry and used for structural color generation.*

## 2.2 Thickness-controlled hybridization and spectral tuning

To identify the physical origin of the transmission-mode response, we first examine how the spectrum evolves across three vertical stack configurations of the same periodic void array. Figures 2a–c compare a periodic array of cylindrical voids of fixed radius $R = 340$ nm and period $P = 1\ \mu m$, while varying only the void depth (see Fig. S1 for additional results at larger radii). In the semi-infinite SRN limit (Fig. 2a), the spectrum exhibits two localized Mie-type modes that define the reference Mie-void response. The major spectral transformation occurs when the same structure is placed in a finite-thickness SRN film (Fig. 2b), where coupling to the film introduces additional resonant features beyond the original localized modes. By contrast, adding the SiO₂ substrate perturbs this already hybridized response only weakly, as reflected by the close similarity between Figs. 2b and 2c. The transmission-mode platform is therefore governed primarily by hybridization between localized Mie-void modes and the finite-film response, while the substrate plays a secondary perturbative role.

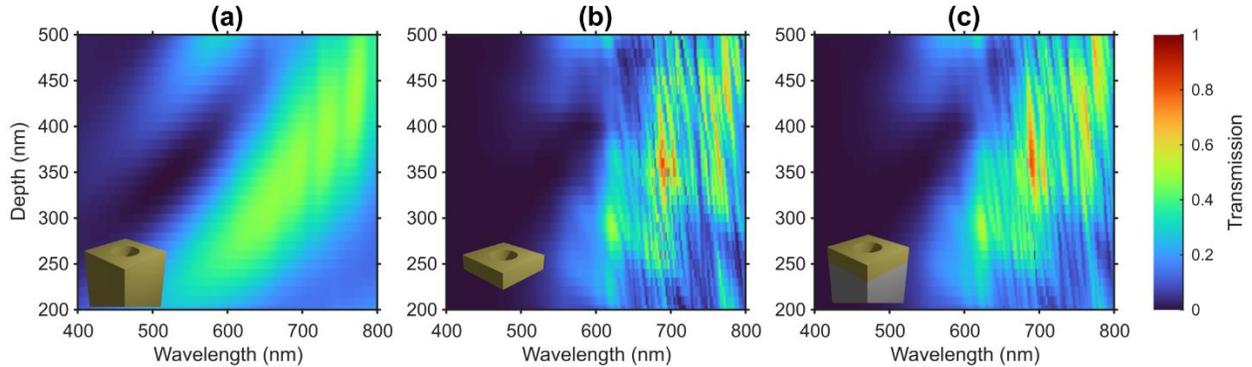

*Figure 2: Simulated transmission spectra for the same cylindrical void geometry across three vertical stack configurations. (a) Semi-infinite SRN medium, showing the reference Mie-void response with localized Mie-type modes. (b) Finite-thickness SRN film, where the original void response evolves through coupling to the finite-film modes. (c) Finite-thickness SRN film on a $SiO_2$ substrate, corresponding to the experimentally relevant configuration. The close similarity between (b) and (c) shows that the dominant spectral transformation arises from introducing the finite SRN film, while the substrate acts mainly as a secondary perturbation of the already hybridized response. The SRN film thickness in (b) and (c) is 580 nm.*

To further clarify the physical origin of the transmission response, we examine the field distributions at the three representative spectral positions marked I–III in Fig. 3a. These wavelengths are not intended to represent isolated resonance extrema, but rather diagnostic points chosen to probe the field composition of the coupled spectrum. The corresponding field maps in Figs. 3b–d show that all three positions retain a clear void-localized contribution, while differing in the extent to which the field is distributed throughout the finite SRN film (the semi-infinite reference case is shown in Fig. S2). Together, these profiles indicate that the transmission architecture supports hybrid modal content combining Mie-void, slab-guided, and Fabry–Pérot-like characteristics. Figure 3 therefore shows that the transmission spectrum is governed not by purely localized resonances, but by their interaction with the finite-film modal environment.

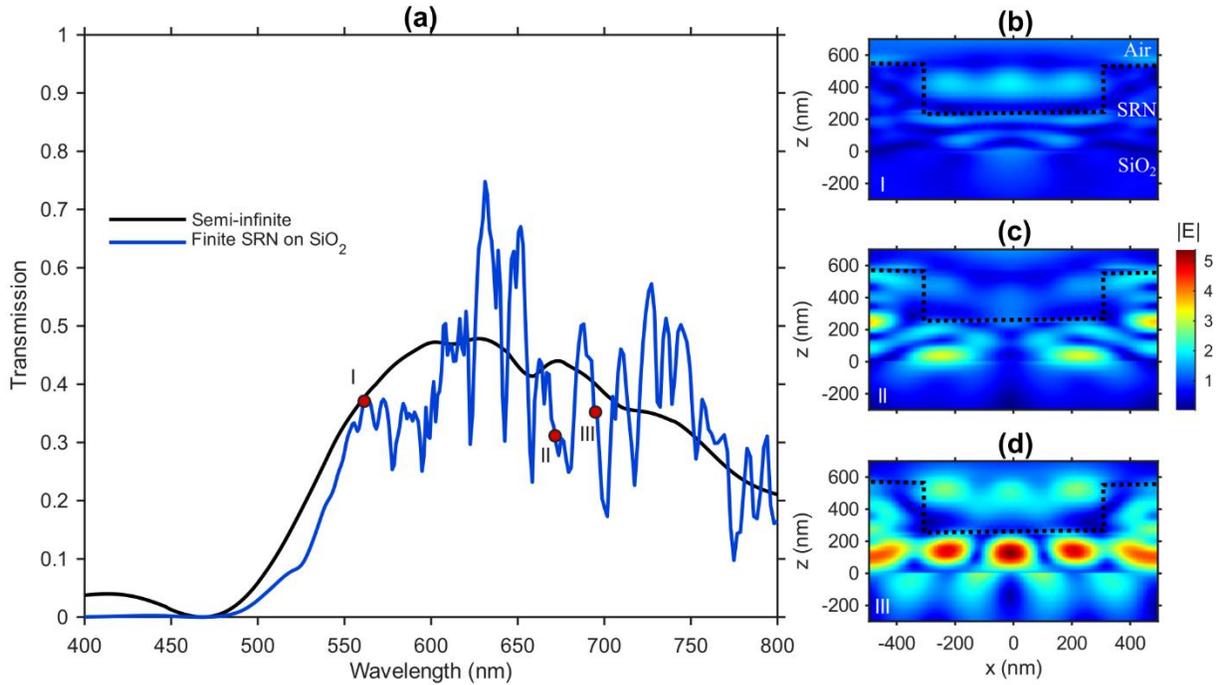

*Figure 3: Modal origin of the transmission resonances in SRN Mie-void structures. (a) Cross-sectional transmission spectra extracted from Fig. 2a,c for a void depth of 340 nm. Transitioning from the semi-infinite SRN medium to a finite-thickness film introduces additional optical modes that couple to the localized Mie-void response. (b–d) Simulated electric-field magnitude distributions, |E|, at λ = 404, 671, and 694 nm, respectively, showing representative modal character across the coupled spectrum. The selected field profiles reveal varying mixtures of void-localized, slab-guided, and Fabry–Pérot-like behavior within the same transmission architecture.*

The thickness-dependent transmission map in Fig. 4a shows that the coupled spectral features evolve along well-defined dispersive branches as the SRN thickness is varied. Within this broader evolution, the circled region highlights a local avoided crossing between two interacting modes. The enlarged view in Fig. 4b makes this behavior explicit: rather than crossing linearly, the two resonances deviate from their uncoupled trajectories and remain separated by a finite spectral gap $\Delta\lambda$ (see inset in Fig. 4b). This avoided crossing is a clear signature of modal coupling, indicating that the interacting states hybridize and exchange character as they approach spectral degeneracy. In the present system, this behavior supports the conclusion that the transmission response is governed by coupling between the underlying Mie-void and finite-film modal families, rather than by independent resonances that shift separately with thickness.

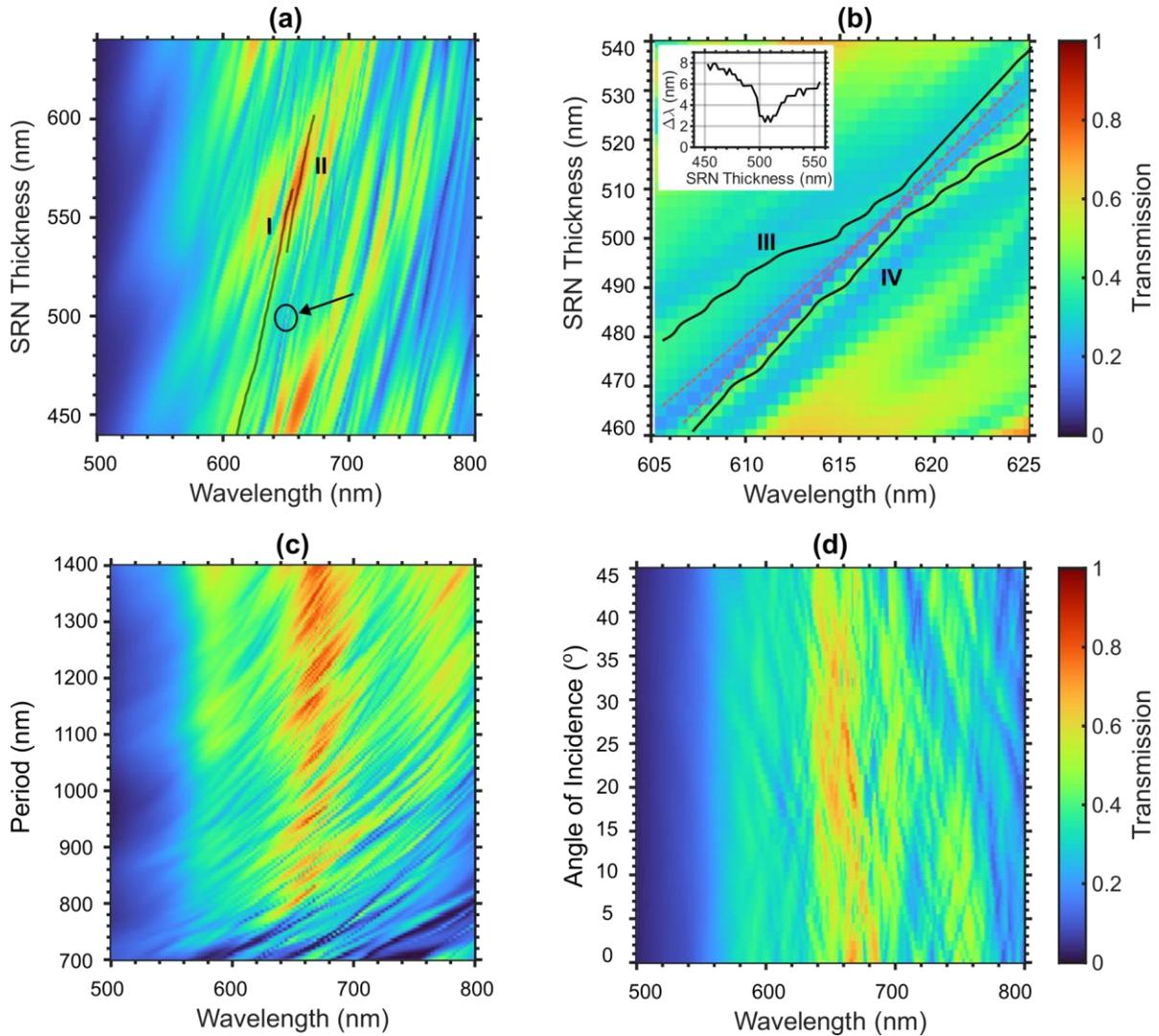

*Figure 4: Dispersion of the coupled resonances in SRN Mie-void metasurfaces. (a) Simulated spectral evolution of the coupled modes as a function of SRN film thickness, showing well-defined dispersive branches across the hybrid transmission regime. Two interacting branches are highlighted. (b) Enlarged view of the region marked in (a), showing a clear avoided crossing between the highlighted modes. The inset indicates the minimum spectral gap between the branches. (c) Simulated spectral evolution as a function of the unit-cell period. Varying the period produces a denser and more strongly restructured spectral landscape, reflecting the additional influence of lateral coupling and diffractive conditions. (d) Simulated spectral evolution as a function of the angle of incidence. Over the examined range, the dominant transmission features remain comparatively stable, particularly near the main transmission band.*

This interpretation is further supported by comparing the influence of thickness, period, and angle of incidence in Figs. 4a,c,d. Varying the SRN thickness produces a clear and orderly evolution of the dominant coupled resonances, including the localized avoided crossing highlighted in Fig. 4b, indicating that the vertical film dimension provides a clean parameter for tracking the hybrid modal energies. By contrast, changing the metasurface period leads to a denser and more strongly restructured spectral landscape (Fig. 4c), consistent with the added influence of lateral coupling and diffractive conditions. The angular response remains comparatively stable over the examined range (Fig. 4d), particularly near the main transmission band. Taken together, these results show that thickness is the clearest parameter for revealing the evolution of the hybrid modes, while period and incidence angle primarily influence how those modes are coupled into and spectrally manifested.

**2.3 Experimental realization of transmission-mode structural colors**

We next validate the transmission-mode platform experimentally by measuring the transmitted spectrum as a function of void depth for fabricated SRN metasurfaces. As shown in Fig. 5a, the measured transmission map exhibits two dominant spectral features whose evolution with depth follows the trends predicted by the simulated coupled spectral response, with one branch shifting through the shorter-wavelength region and a second dispersing at longer wavelengths. Although the experimental spectra are broader and less sharply resolved than their simulated counterparts, the overall modal structure and its depth-dependent evolution are preserved, indicating that the hybrid transmission response remains observable after fabrication. These measurements therefore confirm that void-depth control provides a practical experimental route for accessing and tuning the visible transmission response of the Mie-void metasurface platform.

The spectral evolution in Fig. 5a translates directly into visible color tuning in transmission, as shown by the comparison between measured and simulated coordinates in the CIE 1931 chromaticity diagram (Fig. 5b). As the void depth is varied, the metasurface traces a continuous trajectory in color space, indicating that the coupled transmission resonances produce controlled perceptual changes rather than isolated spectral shifts. The close correspondence between experiment and simulation shows that the observed color response is consistent with the same hybrid spectral picture established above, while the remaining deviations are in line with the spectral broadening and fabrication imperfections evident in the measured transmission map.

Figure 5b therefore connects the depth-dependent resonance evolution to a directly observable transmission-mode color response.

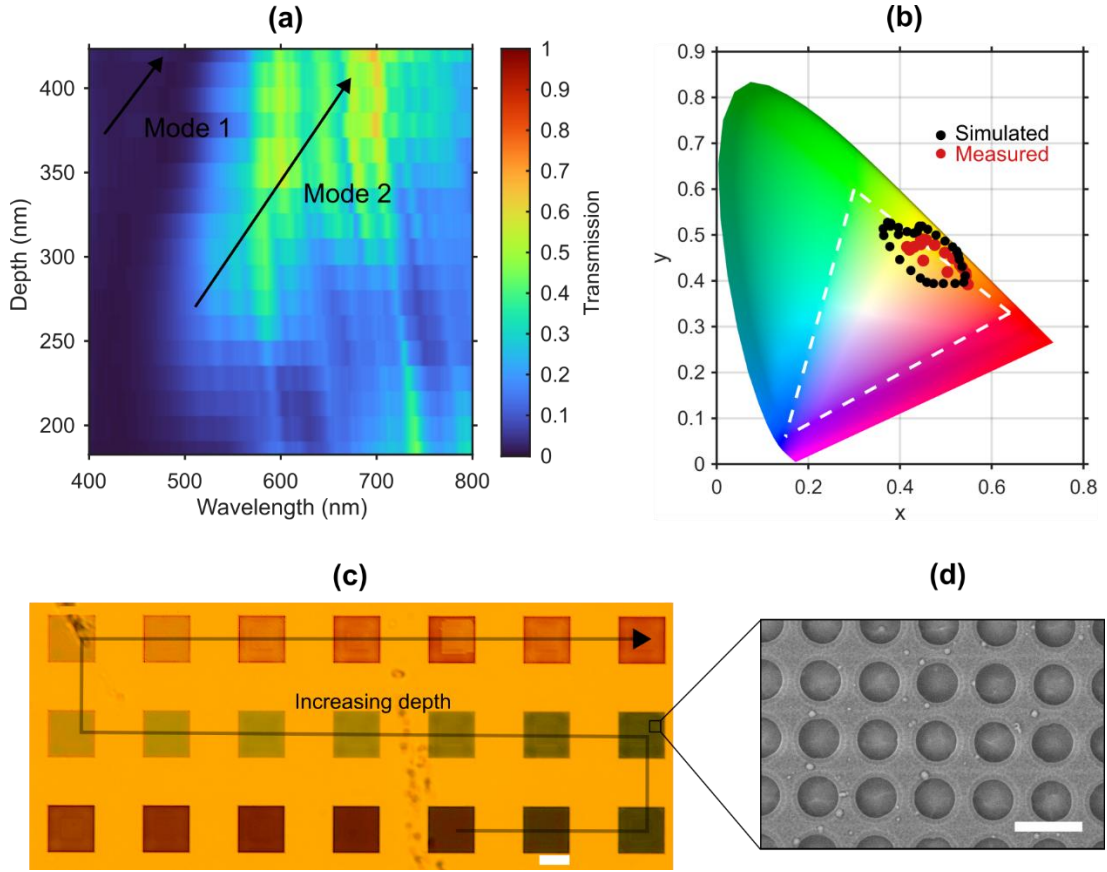

*Figure 5: Experimental validation of the depth-dependent transmission response in SRN Mie-void metasurfaces. (a) Measured transmission spectra as a function of void depth, showing the evolution of the dominant spectral features across the visible range. (b) Corresponding CIE 1931 chromaticity coordinates derived from measured and simulated transmission spectra, showing good agreement between experiment and simulation. (c) Optical microscope image of the metasurface array under white-light transmission. The faded black line indicates the direction of increasing void depth, from 200 to 450 nm. (d) SEM image of a magnified region from (c), confirming uniform formation of the Mie-void array. Scale bar: 1 μm.*

This color tuning is directly visible under white-light transmission, as shown in the optical microscope image of Fig. 5c, where the gradual change in appearance follows the monotonic increase in void depth across the patterned region. The SEM image in Fig. 5d does not resolve the full depth variation across the sample, but it confirms the uniform formation of the Mie-void array within the fabricated metasurface. This supports the interpretation that the measured optical response arises from a well-defined patterned structure rather than from incomplete pattern transfer or local fabrication failure. Taken together, Figs. 5a–d show that the hybrid transmission response predicted numerically is retained in fabricated devices and can be translated into robust visible structural-color tuning.

## 2.4 Spatial encoding with depth-programmed metasurfaces

The depth-dependent transmission response can be extended from uniform test structures to spatially varying metasurfaces by locally modulating the void depth across the patterned area. As shown in Fig. 6a, this approach enables different regions of the same SRN Mie-void metasurface to support distinct transmission states under white-light transmission, producing a spatially resolved color pattern directly in transmission. The resulting contrast is therefore not limited to a single globally tuned resonant condition but can be encoded locally through controlled variation of the metasurface depth profile. This demonstrates that the hybrid transmission response remains addressable on a pixel-by-pixel basis, providing a route toward spatially programmed optical functionality within a single patterned platform.

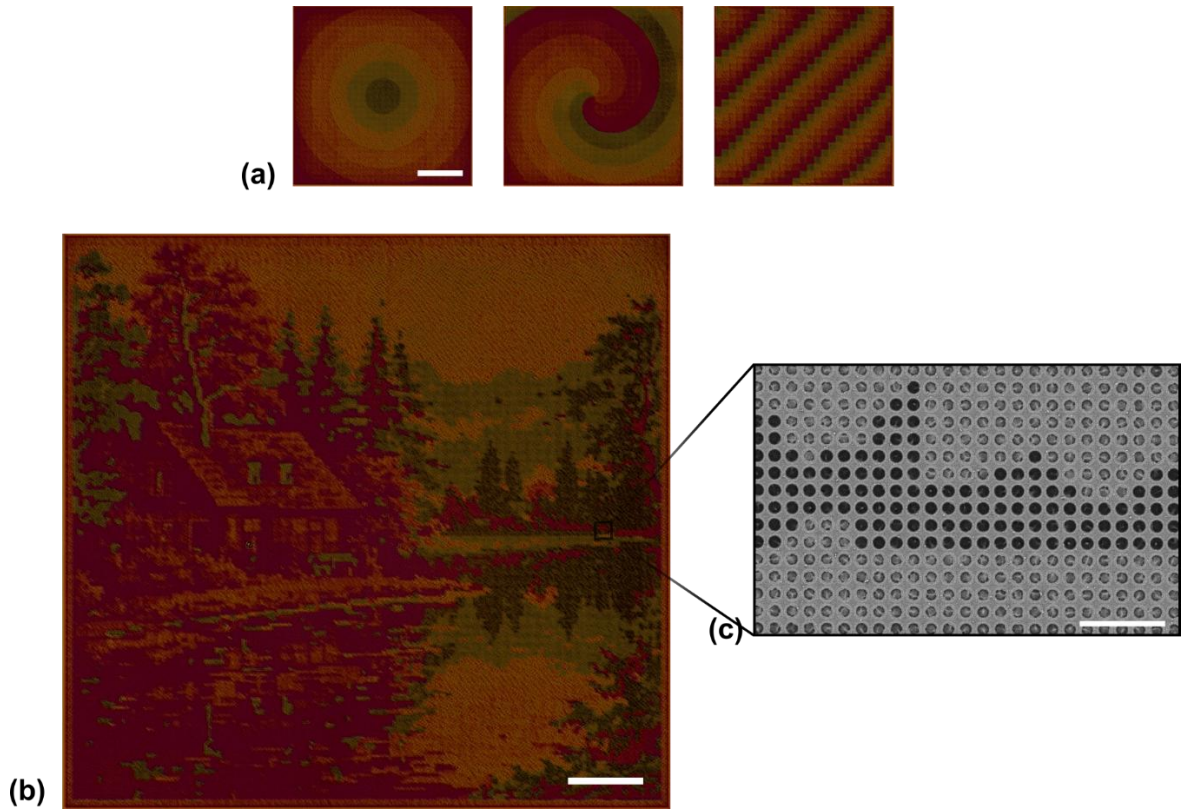

*Figure 6: Spatial encoding of structural-color patterns using depth-programmed SRN Mie void metasurfaces. (a) Optical microscope images in transmission of Mie-void metasurfaces with spatially varying programmed depth, demonstrating controlled generation of simple structural-color patterns. Scale bar: 50 µm. (b) Optical transmission image of a detailed artistic pattern realized using spatial depth modulation of the metasurface. (c) SEM image of an enlarged region from (b), showing local variations in the apparent void filling consistent with the programmed depth modulation used to generate the spatial color pattern. Scale bar: 5 µm.*

This capability is further illustrated in Fig. 6b, where a more intricate transmitted-light pattern is generated by spatially varying the metasurface depth across the design. The enlarged SEM image in Fig. 6c confirms that the observed optical contrast is associated with controlled local modulation of the Mie-void structure. Together, these results show that transmission-mode Mie-void metasurfaces support spatially resolved optical encoding through depth control alone. More broadly, they identify this platform as a viable route toward transmissive optical elements in which spectral response and image formation can be co-designed within the same metasurface architecture.

**Discussion**

The focus of this work is on Mie-void type metasurfaces operating in transmission. Such elements offer distinct advantages over their reflection counterpart, e.g. they allow for straight-line optical paths, making them easier to integrate into compact, linear systems like microscopes or fiber-coupled setups. This could be important in display applications, where backlight is used to illuminate the device, and the Mie-void type metasurface is used as a selective bandpass filter, choosing the desired color. A central outcome of this work is that the Mie-void design paradigm remains physically meaningful after the transition from reflection to transmission. Although the finite-film transmission geometry introduces a more complex optical environment than the semi-infinite reflection limit, the response continues to retain a clear void-localized resonant contribution. The core logic of Mie-void metasurfaces is therefore not lost in the visible transmission regime and it is preserved in a modified form that remains both spectrally tunable and experimentally accessible. Rather than replacing the inverse-resonator concept, transmission extends it into a more elaborate yet still designable photonic setting.

From the physics point of view, what is new in transmission is that the Mie-void response no longer appears as an isolated localized resonance, but as part of a coupled finite-film system. The finite SRN layer introduces slab-guided and Fabry–Pérot-like modal contributions that hybridize with the underlying void resonance, producing a richer spectral structure than in the reflection-oriented limit. This hybridization is seen most clearly in the thickness-dependent dispersion and the avoided crossing observed in the coupled-mode evolution, which indicate that the relevant transmission states are mixed rather than independent. In this context, film thickness provides the clearest parameter for revealing how the hybrid modal energies evolve, while lateral periodicity and incidence angle primarily influence how the coupled spectrum is accessed and expressed. The transmission platform should therefore be understood not as a simple spectral translation of reflection-mode Mie-voids, but as a distinct operating regime with its own modal behavior and design considerations.

Taken together, these results extend Mie-void metasurfaces beyond reflective operation and identify them as a viable platform for transmissive optical functionality in the visible. Because the transmission response can be tuned spectrally through depth control and encoded spatially across a patterned surface, the platform is naturally relevant to compact transmissive elements such as spectral filters, image-encoding structures, and display-oriented components. More broadly, the present work shows that inverse-dielectric resonator concepts can remain effective even in finite-film environments where additional modal channels are unavoidable. In this sense, transmission-mode Mie-void metasurfaces represent not just a new implementation of a known resonant geometry, but an extension of the inverse-dielectric metasurface concept into a practically useful and physically distinct operating regime.

## Methods

*Sample preparation*

A thin silicon-rich nitride (SRN) film (~580 nm) was deposited on a bare 15 × 15 mm glass substrate (Borofloat 33) by plasma-enhanced chemical vapor deposition (PECVD; PlasmaLab 100). Additional details of the SRN deposition are provided in Ref[44]. The sample was prebaked on a hot plate at 180°C for 5 min, followed by spin-coating of a positive electron-beam resist (PMMA 495A) at 2300 rpm for 50 s and a subsequent bake at 180°C for 2 min. A second PMMA layer was deposited using the same procedure, yielding a total resist thickness of approximately 800 nm.

Patterning was performed by electron-beam lithography (Elionix ELS-G100) using a beam current of 1 nA. For the arrayed Mie-void metasurfaces, the different exposure doses were assigned using an internal dose-modulation function of the electron-beam lithography software. For the spatially varying structures, BEAMER software (GenISys) was used to distribute the exposure dose according to the target design. Proximity-effect correction was applied to compensate for unintended spatial variations in the delivered dose.

After exposure, the sample was developed in an MIBK:IPA solution for 1 min, rinsed in IPA for 10 s, and dried under $N_2$. Pattern transfer into the SRN film was then carried out by reactive ion etching (Corial 2001) using an $SF_6/CH_4$ recipe for 690 s, corresponding to an etch rate of approximately 0.7 nm/s.

*Optical measurements*

Transmission spectra were measured using a Nikon Eclipse TE300 inverted microscope illuminated by a 100 W tungsten-halogen lamp. The transmitted light was collected with a 50× objective and imaged onto an intermediate plane, where a field stop selected the region of interest. The light was then coupled into a spectrometer using a 50 mm achromatic doublet lens (Thorlabs AC254-050-AB) and a multimode fiber (Ocean Insight QP600-2-VIS-NIR).

Optical microscope images were acquired using an Olympus BX53M upright microscope with MPLFLN-BD 20× and 50× objectives.

*Simulations*

Numerical simulations were performed using commercially available rigorous coupled-wave analysis (RCWA) software (Ansys Lumerical). Three vertical stack configurations were considered:

(i) a semi-infinite SRN medium, with air above and SRN extending below the patterned region;
(ii) a finite-thickness SRN film suspended in air; and

(iii) a finite-thickness SRN film supported on a $SiO_2$ substrate.

Unless otherwise stated, all simulations considered cylindrical etched regions with radius $R = 340$ nm and depths ranging from 200 to 500 nm, defining the voids within the SRN film. The unit-cell period was 1 $\mu$m. The optical constants of SRN were taken from Ref[10] ($\beta = 2.99$). Electric-field distributions were extracted from a field monitor placed at $y = 0$. Transmission spectra were obtained from the grating-power output, and only the zeroth-order transmitted component was analyzed.

**References**

1   Kamali SM, Arbabi E, Arbabi A, Faraon A. A review of dielectric optical metasurfaces for wavefront control. *Nanophotonics* 2018; **7**: 1041–1068.

2   Khorasaninejad M, Capasso F. Metalenses: Versatile multifunctional photonic components. *Science (1979)* 2017; **358**. doi:10.1126/science.aam8100.

3   Yu N, Capasso F. Flat optics with designer metasurfaces. *Nat Mater* 2014; **13**: 139–150.

4   Zou X, Zhang Y, Lin R, Gong G, Wang S, Zhu S et al. Pixel-level Bayer-type colour router based on metasurfaces. *Nat Commun* 2022; **13**: 3288.

5   Yang W, Xiao S, Song Q, Liu Y, Wu Y, Wang S et al. All-dielectric metasurface for high-performance structural color. *Nat Commun* 2020; **11**. doi:10.1038/s41467-020-15773-0.

6   Yang C, Wang Z, Yuan H, Li K, Zheng X, Mu W et al. All-Dielectric Metasurface for Highly Tunable, Narrowband Notch Filtering. *IEEE Photonics J* 2019; **11**: 1–6.

7   Khorasaninejad M, Chen WT, Devlin RC, Oh J, Zhu AY, Capasso F. Metalenses at visible wavelengths: Diffraction-limited focusing and subwavelength resolution imaging. 2016https://www.science.org.

8   Devlin RC, Khorasaninejad M, Chen WT, Oh J, Capasso F. Broadband high-efficiency dielectric metasurfaces for the visible spectrum. *Proc Natl Acad Sci U S A* 2016; **113**: 10473–10478.

9   Engelberg J, Zhou C, Mazurski N, Bar-David J, Kristensen A, Levy U. Near-IR wide-field-of-view Huygens metalens for outdoor imaging applications. *Nanophotonics* 2020; **9**: 361–370.

10  Goldberg O, Mazurski N, Levy U. Silicon rich nitride: a platform for controllable structural colors. *Nanophotonics* 2024. doi:10.1515/nanoph-2024-0454.

11  Proust J, Bedu F, Gallas B, Ozerov I, Bonod N. All-Dielectric Colored Metasurfaces with Silicon Mie Resonators. *ACS Nano* 2016; **10**: 7761–7767.

12  Yang J-H, Babicheva VE, Yu M-W, Lu T-C, Lin T-R, Chen K-P. Structural Colors Enabled by Lattice Resonance on Silicon Nitride Metasurfaces. *ACS Nano* 2020; **14**: 5678–5685.

13  Han Z, Frydendahl C, Mazurski N, Levy U. MEMS cantilever–controlled plasmonic colors for sustainable optical displays. *Sci Adv* 2022; **8**: 889.

14  Zhu X, Yan W, Levy U, Mortensen † N Asger, Kristensen A. Resonant laser printing of structural colors on high-index dielectric metasurfaces. doi:10.1126/sciadv.1602487.

15  Arbabi A, Arbabi E, Kamali SM, Horie Y, Han S, Faraon A. Miniature optical planar camera based on a wide-angle metasurface doublet corrected for monochromatic aberrations. *Nat Commun* 2016; **7**: 13682.

16  Decker M, Staude I, Falkner M, Dominguez J, Neshev DN, Brener I *et al.* High-Efficiency Dielectric Huygens' Surfaces. *Adv Opt Mater* 2015; **3**: 813–820.

17  Zhou Y, Kravchenko II, Wang H, Nolen JR, Gu G, Valentine J. Multilayer Noninteracting Dielectric Metasurfaces for Multiwavelength Metaoptics. *Nano Lett* 2018; **18**: 7529–7537.

18  Arbabi A, Horie Y, Bagheri M, Faraon A. Dielectric metasurfaces for complete control of phase and polarization with subwavelength spatial resolution and high transmission. *Nat Nanotechnol* 2015; **10**: 937–943.

19  Wu Z, Zhang Z, Xu Y, Zhai Y, Zhang C, Wang B *et al.* Random color filters based on an all-dielectric metasurface for compact hyperspectral imaging. *Opt Lett* 2022; **47**: 4548.

20  Tittl A, Leitis A, Liu M, Yesilkoy F, Choi D-Y, Neshev DN *et al.* Imaging-based molecular barcoding with pixelated dielectric metasurfaces. *Science (1979)* 2018; **360**: 1105–1109.

21  Lee J, Park Y, Kim H, Yoon Y, Ko W, Bae K *et al.* Compact meta-spectral image sensor for mobile applications. *Nanophotonics* 2022; **11**: 2563–2569.

22  Fu R, Chen K, Li Z, Yu S, Zheng G. Metasurface-based nanoprinting: principle, design and advances. *Opto-Electronic Science* 2022; **1**: 220011.

23  Kim I, Jang J, Kim G, Lee J, Badloe T, Mun J *et al.* Pixelated bifunctional metasurface-driven dynamic vectorial holographic color prints for photonic security platform. *Nat Commun* 2021; **12**: 3614.

24  Wang D, Li Y-L, Zheng X-R, Ji R-N, Xie X, Song K *et al.* Decimeter-depth and polarization addressable color 3D meta-holography. *Nat Commun* 2024; **15**: 8242.

25   Gong J, Xiong L, Pu M, Li X, Ma X, Luo X. Visible Meta-Displays for Anti-Counterfeiting with Printable Dielectric Metasurfaces. *Advanced Science* 2024; **11**. doi:10.1002/advs.202308687.

26   Tian Z, Zhu X, Surman PA, Chen Z, Sun XW. An achromatic metasurface waveguide for augmented reality displays. *Light Sci Appl* 2025; **14**: 94.

27   Lee G-Y, Hong J-Y, Hwang S, Moon S, Kang H, Jeon S *et al.* Metasurface eyepiece for augmented reality. *Nat Commun* 2018; **9**: 4562.

28   zhang Q, Liu C, Gan G, Cui X. Visible perfect reflectors realized with all-dielectric metasurface. *Opt Commun* 2017; **402**: 226–230.

29   Hsiao H, Chu CH, Tsai DP. Fundamentals and Applications of Metasurfaces. *Small Methods* 2017; **1**. doi:10.1002/smtd.201600064.

30   Overvig AC, Shrestha S, Malek SC, Lu M, Stein A, Zheng C *et al.* Dielectric metasurfaces for complete and independent control of the optical amplitude and phase. *Light Sci Appl* 2019; **8**. doi:10.1038/s41377-019-0201-7.

31   Chen Z, Mazurski N, Engelberg J, Levy U. Tunable Transmissive Metasurface Based on Thin-Film Lithium Niobate. *ACS Photonics* 2025; **12**: 1174–1183.

32   Hentschel M, Koshelev K, Sterl F, Both S, Karst J, Shamsafar L *et al.* Dielectric Mie voids: confining light in air. *Light Sci Appl* 2023; **12**. doi:10.1038/s41377-022-01015-z.

33   Arslan S, Kappel M, Canós Valero A, Tran TTH, Karst J, Christ P *et al.* Attoliter Mie Void Sensing. *ACS Photonics* 2025; **12**: 3950–3958.

34   Zheng Z, Smirnova D, Sanderson G, Cuifeng Y, Koutsogeorgis DC, Huang L *et al.* Broadband infrared imaging governed by guided-mode resonance in dielectric metasurfaces. *Light Sci Appl* 2024; **13**: 249.

35   Huang L, Jin R, Zhou C, Li G, Xu L, Overvig A *et al.* Ultrahigh-Q guided mode resonances in an All-dielectric metasurface. *Nat Commun* 2023; **14**: 3433.

36   Ding L, Wang KJ, Wang W, Zhu DF, Yin CY, Liu JS. Experimental verification and investigation of disks scattering slab modes in metal-dielectric heterostructures. *Sci Rep* 2013; **3**: 2493.

37   Danaeifar M, Granpayeh N. Analytical synthesis of high-Q bilayer all-dielectric metasurfaces with coupled resonance modes. *Appl Opt* 2022; **61**: 338.

38   Liu W, Li Z, Cheng H, Chen S. Dielectric Resonance-Based Optical Metasurfaces: From Fundamentals to Applications. *iScience* 2020; **23**: 101868.

39   Berkhout A, Wolterink TAW, Koenderink AF. Strong Coupling to Generate Complex Birefringence: Metasurface in the Middle Etalons. *ACS Photonics* 2020; **7**: 2799–2806.

40   Mansha S, Tapar J, Kuznetsov AI, Paniagua-Domínguez R. Anti-crossing of modes and singularity points in dielectric metasurface-on-mirror microcavities. *Opt Express* 2025; **33**: 47273.

41   Park C-S, Koirala I, Gao S, Shrestha VR, Lee S-S, Choi D-Y. Structural color filters based on an all-dielectric metasurface exploiting silicon-rich silicon nitride nanodisks. *Opt Express* 2019; **27**: 667.

42   Upadhyayula KK, Wardenberg L, Schilling J. Enhancement of second-harmonic generation in silicon-rich nitride using photonic bound states in the continuum. *Opt Express* 2025; **33**: 24957.

43   Goldberg O, Reisinger A, Mazurski N, Magdassi S, Levy U. NIR Imaging in the Visible via all Optical Dynamic Dielectric Metasurface. In: *CLEO 2025*. Optica Publishing Group: Washington, D.C., 2025, p FF125_1.

44   Goldberg O, Gherabli R, Engelberg J, Nijem J, Mazurski N, Levy U. Silicon Rich Nitride Huygens Metasurfaces in the Visible Regime. *Adv Opt Mater* 2024; **12**. doi:10.1002/adom.202301612.

45   Goldberg O, Mazurski N, Levy U. Single-Step Grayscale Lithography of Multi-Depth Mie Void Metasurfaces. 2026.